\documentclass[twocolumn,showpacs,amsmath,amssymb,prl]{revtex4}
\usepackage{graphicx}
\usepackage{dcolumn}
\usepackage{bm}

\begin{document}

\title{A Primer for Electro-Weak Induced Low Energy Nuclear Reactions}

\author{Y.N. Srivastava}
\affiliation{Dipartimento di Fisica \& INFN, Universita`degli Studi di Perugia, 
06123 Perugia, Italy}
\author{A. Widom}
\affiliation{Physics Department, Northeastern University, Boston MA 02115, U.S.A}
\author{L. Larsen}
\affiliation{Lattice Energy LLC, 175 North Harbor Drive, Chicago IL 60601, U.S.A.}

\begin{abstract}
In a series of papers, cited in the main body of the paper below, detailed calculations have 
been presented which show that electromagnetic and weak interactions can induce low energy 
nuclear reactions to occur with observable rates for a variety of processes. A common element 
in all these applications is that the electromagnetic energy stored in many relatively slow 
moving electrons can -under appropriate circumstances- be collectively transferred into fewer, 
much faster electrons with energies sufficient for the latter to combine with protons (or deuterons, 
if present) to produce neutrons via weak interactions. The produced neutrons can then initiate 
low energy nuclear reactions through further nuclear transmutations. The aim of this paper is 
to extend and enlarge upon various examples analyzed previously, present simplified order of 
magnitude estimates for each and to illuminate a common unifying theme
amongst all of them.
 
\end{abstract}


\maketitle

\section{1. Introduction}

In a set of papers\cite{LW1,LW2,LW3,WSL1,WSL2,WSL3}, diverse physical processes 
have been considered and detailed computations on them performed to reveal 
conditions under which, in each case, the end result is a low energy 
nuclear reaction (LENRs) {\it induced by electro-weak interactions}. Even 
though weak interactions are an integral part 
of the Standard Model of fundamental interactions, which unifies the electromagnetic 
force with the weak and the strong (nuclear) force, nonetheless low energy applications 
of weak interactions in condensed matter devices are novel and hence unfamiliar. 
Practically all existing condensed matter devices are essentially of electromagnetic 
origin. There are sound reasons for the latter circumstance. Charged particles in 
condensed matter (electrons or ions) normally possess low kinetic energies (typically 
a few eV or less) and yet they can trigger substantial electromagnetic processes which 
one can usefully harness. In sharp contrast, for an electron to undergo a weak interaction 
say with a proton in an ion and produce a neutron requires MeV range of energies (due the 
fact that the neutron is heavier than the proton by about \begin{math}1.3\ MeV\end{math} 
and hence there is an energy ``threshold'' which must be overcome). It follows then that 
for neutron production (and subsequent nuclear transmutations) via weak interactions, 
special conditions in condensed matter systems must be found which accelerate an electron 
to MeV range of energies. Successful avenues to accomplish precisely this purpose have been 
described in the papers quoted above. The present paper is devoted to delineating the unifying 
features and to an overall synthesis of these rather different collective processes.

In\cite{LW1}, metallic hydride surfaces on which plasma oscillations exist were analyzed. It 
was shown that the collective plasma oscillations on the surface can contribute some of their 
electric energy to an electron so that the following reaction becomes kinematically allowed
\begin{equation}
\label{W0}
W_{electric} + e^- + p \rightarrow\ n + \nu_e.
\end{equation} 
The relevant scale of the electric field \begin{math}{\cal E}\end{math} and the plasma frequency 
\begin{math}\Omega\end{math} needed to accelerate electrons to trigger neutron production is found to be
\begin{equation}
\label{W1}
\frac{c {\cal E}}{\Omega} = (\frac{mc^2}{e})\ \approx\ 0.5\ \times 10^6\ Volts,
\end{equation}
where \begin{math}c\end{math} is the speed of light, \begin{math}m\end{math} is the mass and 
\begin{math}(-e)\end{math} the charge of the electron.

The particular condensed matter environment leads in this case to ultra-cold (that is, ultra  
low-momentum) neutrons. These neutrons born ultra-cold have extraordinarily large nuclear absorption 
cross-sections and thus a high probability of producing nuclear transmutations and an extremely 
low probability of neutrons escaping beyond micron-scale and smaller surface regions in which
they are formed. There is also a high suppression 
in the production of high energy gamma rays\cite{LW2}. For such metallic chemical cells, 
comprehensive calculations of the rates of LENRs\cite{LW3} were made which confirmed a robust 
production of new elements.
   
In\cite{WSL1}, a magnetic analog of the above (the so-called exploding wire problem) was analyzed. 
We found that a strong electric current carrying wire\cite{wire} can -under suitable conditions- 
channel the collective magnetic energy sufficiently once again to excite a certain fraction of 
electrons to undergo the weak interaction process
\begin{equation}
\label{W2}
W_{magnetic} + e^- + p \rightarrow\ n + \nu_e.
\end{equation}
The scale of current required here was shown to be of the order of the Alfv$\acute{e}$n current 
\begin{math}I_o \end{math}  
\begin{equation}
\label{W3}
I_o\ =\ (\frac{mc^2}{e})(\frac{4\pi}{R_{vac}})\ \approx\  17\ KiloAmperes
\end{equation}
Observation of copious neutrons in exploding wire experiments is by now legion
\cite{exp1,exp2,exp3,exp4,exp5,exp6}. Experimentally, neutron production has also been confirmed  
for lightning\cite{lightning}, the big exploding wire in the sky, where typical currents are about 
\begin{math}30\end{math} KiloAmperes and higher.   

Quite recently\cite{WSL2}, another application of the magnetic mode inducing LENR has been made 
to unravel the mystery surrounding observed particle production and nuclear transmutations in the 
Solar corona and in Solar flares\cite{C1,C2,C3,C4,C5,C6,C7,C8,C9}. Spectacular pictures of flux 
tubes are now available\cite{swede} showing giant magnetic flux tubes exiting out of one sunspot 
and entering into another. We show theoretically\cite{WSL2} how these can lead to steady LENR. In 
fierce solar flares, we find that as a flux tube disintegrates it generates electric fields strong 
enough to accelerate electrons and protons toward each other with a center of mass energy of 
~\begin{math} 300\ GeV \end{math}, equivalent to the highest energy electron-proton colliding beam 
(HERA) built on Earth. For a strong solar flare which occurred on July 14, 2000, we computed the 
flux of muons which reach our Earth. Our theoretical flux agrees quite nicely with the experimental 
data by the L3+C Collaboration at LEP through their observation of high energy muons produced in 
coincidence with this huge flare\cite{C9}.  

In the present paper, we shall try to provide a unified picture of electro-weak (EW) induced LENRs 
bringing out the essential physics and omitting many technical details all of which can be found in our 
earlier papers. Order of magnitude estimates of relevant parameters for different physical processes 
utilizing the electric and/or the magnetic modes  shall be presented to stress the feasibility of LENRs. 
The paper is organized as follows. In sec.2, general considerations revealing the basic idea behind 
EW induction of LENR is given. In sec.3, the case of metallic hydride cells where electric charge 
fluctuations play a major role is discussed along with estimates of expected rates of nuclear 
transmutations. Estimates of the mean free path for the ultra low momentum neutrons and MeV range gamma 
rays are shown to be so short as to confine them to the material, i.e. they get absorbed on the surface.
In sec.4, we consider strong electric currents flowing through thin wires to show how the collective 
magnetic mode energy generates a huge chemical potential in the MeV range sufficient to induce LENRs. 
In sec.5, applying the mechanism presented in sec.4 we show how a giant transformer -or a betatron- 
is generated for the solar corona leading to LENRs and, for solar flares the production of extremely 
high energy particles exposing the rich structure of the Standard Model. Sec.6 closes the paper with 
a summary of our results along with some concluding remarks and future outlook.

\section{2. Genesis of EW Induced LENR}       
 
A free neutron can and does decay via weak interaction into a proton, an electron and an \begin{math}\bar {\nu}_e\end{math} 
\begin{equation}
\label{EW0}
n \rightarrow\ p + e^- + \bar{\nu}_e, 
\end{equation}
since the Q-value for this reaction, \begin{math} Q\ =\ (M_n - M_p - m)c^2\ \approx\ 0.78\ MeV\end{math} is positive. On the other hand, the production of a neutron through the inverse reaction \begin{math} e^-\ +\ p\ \rightarrow\ n +\ \nu_e\end{math} for electrons and protons of very low kinetic energy (generally to be found in condensed matter systems) is kinematically forbidden unless the energy of the incident (ep) system can be augmented by this Q-value. Hence, to induce  LENR through weak interaction such as 
\begin{equation}
\label{EW1}
W\ + e^- + p \rightarrow\ n + \nu_e,
\end{equation}
to be kinematically allowed, we require an energy  \begin{math}W\ \geq\ 0.78\ MeV \end{math} to be fed in to the
\begin{math}(e^- p)\end{math} -system originally with negligible kinetic energy.
Thus our first task is to find a mechanism within a condensed matter system which can supply MeV scale energies to accelerate an electron to overcome the threshold barrier. Since electrons are accelerated by an electric field through the equation \begin{math}\dot{p}\ =\ (-e) {\bf E}\end{math}, let us assume that on a metal surface a sinusoidal electric field exists: \begin{math} {\bf E}\ =\ {\bf E}_o cos(\Omega t)\end{math} of frequency \begin{math}\Omega\end{math}. The average change in the momentum (\begin{math}\Delta p\end{math}) is then easily obtained through \begin{math}(\Delta p)^2\ = (e^2  \bar{E^2}/\Omega^2)\end{math}. The average (squared) total energy \begin{math}\bar{K^2}\end{math} for an electron of rest mass \begin{math}m\end{math} with an original small momentum \begin{math}{\bf p}\end{math} is given by
\begin{equation}
\label{EW2}
\bar{K^2} = (mc^2)^2 + (c{\bf p})^2 + \frac{e^2c^2\bar{E^2}}{\Omega^2} = (c{\bf p})^2 + (mc^2)^2[1 +  \frac{\bar{E^2}}{{\cal E}^2}],
\end{equation}    
where the relevant scale of the required electric field for neutron production is set by \begin{math}{\cal E}\ =\ (mc/\hbar)(\hbar\Omega/e)\end{math}\cite{mom}. For metallic hydride surfaces upon which plasma oscillations exist, typical values for the surface plasmon polariton frequencies are in the range \begin{math}(\hbar\Omega/e) \approx\ (5-6) \times 10^{-2}\end{math} Volts, thereby requiring \begin{math}{\cal E}\ \approx\ 2 \times 10^{11} Volts/meter\end{math}. To put it in perspective,  
let us recall that typical atomic electric fields are of this order of magnitude. For consider, the electric field when located a Bohr radius (\begin{math}a\ \approx\ 0.5\ Angstrom\end{math}) away from an isolated proton. It is given by: \ \begin{math}{\cal E}(a)\ =\ (e/a^2)\ \approx\ 5\times 10^{11} Volts/meter\end{math}. Coherent proton oscillations on a metallic hydride mono layer will be shown in the next section to produce electric fields and plasma frequencies of the order of magnitudes needed for neutron production. There we shall show that neutrons are born with very small momentum (ultra cold) because their production is collective through a large number of protons coherently oscillating over a macroscopic region of the mono layer surface. Since the nuclear absorption cross-section of ultra low momentum neutrons is extremely large, it has two desirous effects:(i) the nuclear transmutation probability is large which makes the rates substantial and (ii) the mean free path for a neutron to escape outside the metal surface is reduced to atomic distances. Hence no free neutrons for this process. There is in addition a photon shield created 
by mass-renormalized electrons which inhibits MeV range gamma rays to escapeing the surface region. We shall review it in the next section.    

Let us now turn to the magnetic mode of exciting ``electron capture by proton'' in a strong current carrying wire in the steady state. For a wire of length \begin{math}\Lambda\end{math} carrying a steady current \begin{math}I\end{math} with N flowing electrons, the collective kinetic energy due to the motion of all the electrons  is most simply described through the inductive energy formula
\begin{equation}
\label{EW5}
W = (1/2c^2) L I^2,
\end{equation}
where \begin{math}L\ =\ \eta\ \Lambda\end{math}, is the inductance and \begin{math}\eta\end{math} (of order unity) is the inductance per unit length. If an electron is removed (by any means, such as when an electron is destroyed in a weak interaction), the change in the current is 
\begin{equation}
\label{EW6}
\delta I = -e (\frac{v}{\Lambda}),
\end{equation}
where \begin{math}v\end{math} is the mean velocity of the electrons in the current. The chemical potential is then
given by
\begin{equation}
\label{EW7}
\mu = - \frac{\partial W}{\partial N} = -(\frac{L}{c^2}) I[I (N) - I(N - 1)] = \frac{e \eta I v}{c^2}.  
\end{equation}
We may write it in a more useful (system of unit independent) form using the Alfv$\acute{e}$n current \begin{math}I_o\ \approx\ 17 KiloAmperes\end{math}, which has been defined in Eq.(\ref{W3}). 
\begin{equation}
\label{EW8}
\mu = (mc^2)\eta(\frac{I}{I_o})(\frac{v}{c}). 
\end{equation}
Thus, we see that even with a moderate \begin{math}(v/c)\ \approx\ 0.1\end{math}, if currents are much larger than the Alfv$\acute{e}$n value, the chemical potential can be of the order of MeV's or higher. This is an example of how the collective magnetic kinetic energy can be distributed to accelerate a smaller number of particles with sufficient energy to produce neutrons. Further discussions about the exploding wires are postponed until sec.4.

Let us now consider the solar corona for which the magnetic flux geometry is different from that of a wire. In a wire, the magnetic field loops surround the flowing current. In the corona, there are oppositely directed currents of electrons and protons which loop around the walls of magnetic flux tubes. In a steady state magnetic flux tube which enters the solar corona out of one sun spot and returns into another sun spot without exploding, there is a substantial amount of stored magnetic energy. For a small change \begin{math}\delta I\end{math}, in the current going around the vortex circumference, the change in the magnetic energy \begin{math}\delta {\cal E}_{mag}\end{math} is given by
\begin{equation}
\label{EW9}
\delta {\cal E}_{mag} = (\frac{1}{c}) \delta \Phi I.
\end{equation}
If the length \begin{math}L\end{math} denotes the vortex circumference, then -as described previously- for the destruction of an electron in the weak interaction, the change in the current corresponds to
\begin{equation}
\label{EW10}
\delta I = -e (\frac{v}{L}),
\end{equation} 
where \begin{math}v\end{math} denotes the relative tangential velocity between the electron and the proton. Setting
\begin{math}\Phi\ =\ B \Delta S\end{math} and \begin{math}\delta {\cal E}_{mag}\ =\ - W_{mag}\end{math}, we obtain
\begin{equation}
\label{EW11}
W_{mag} = (eB)(\frac{\Delta S}{L})(\frac{v}{c}). 
\end{equation}
For a cylindrical flux tube, 
\begin{equation}
\label{EW12}
(\frac{\Delta S}{L}) = \frac{\pi R^2}{2 \pi R} = \frac{R}{2}. 
\end{equation}
For numerical estimates for the sun, it is useful to rewrite the above as
\begin{equation}
\label{EW13}
W_{mag} \approx\ (15\ GeV) (\frac{B}{Kilogauss})(\frac{R}{Kilometers})(\frac{v}{c}). 
\end{equation}
For an estimate, consider some typical values
\begin{eqnarray}
\label{EW14}
R\ \approx\ 10^2 Kilometers\cr
B\ \approx\ 1 Kilogauss\cr
\cr
\frac{v}{c}\ \approx\ 10^{-2}\cr
\cr
W_{mag}\ \approx\ 15\ GeV. 
\end{eqnarray}
Thus, even when the flux tube does not explode, appreciable neutron production is to be expected. It should be noted that
neutrons produced via weak interactions in the higher-energy regime dominated by collective magnetic effects do not necessarily
have ultra low-momentum.   

On the other hand, for a spectacular solar fare which lasts for a time \begin{math}\Delta t\end{math}, the loss of magnetic flux tube would yield a mean Faraday law acceleration voltage \begin{math}\bar{V}\end{math} around the walls 
given by
\begin{equation}
\label{EW15}
\bar{V} = \frac{\Delta \Phi}{c\Delta t}.
\end{equation}

Inserting as before \begin{math}\Delta \Phi\ =\ B \Delta S\end{math}, where \begin{math}B\end{math} denotes the mean magnetic field before the explosion and \begin{math}\Delta S\end{math} the inner cross-sectional area of the flux tube, we have for the mean acceleration energy
\begin{equation}
\label{EW16}
e\bar{V} = (eB)\frac{\Delta S}{\Lambda}, where\ \Lambda\ =\ c \Delta t
\end{equation}
For a cylindrical geometry, we may again rewrite it in a useful form
\begin{equation}
\label{EW17}
e \bar{V} \approx\ (30\ GeV) (\frac{B}{Kilogauss})(\frac{\pi R^2}{\Lambda -Kilometers}). 
\end{equation}
For a coronal mass ejecting coil exploding in a time \begin{math}\Delta t\ \approx\ 10^2\end{math} seconds, we may
estimate
\begin{eqnarray}
\label{EW18}
R\ \approx\ 10^4 Kilometers\cr
B\ \approx\ 1 Kilogauss\cr
\Lambda\ \approx\ 3 \times\ 10^7\ Kilometers\cr
e \bar{V}\ \approx\ 300\ GeV. 
\end{eqnarray}
Physically, it corresponds to a colliding beam of electrons and protons with a center of mass energy of \begin{math}300\ GeV\end{math}. More on these matters in sec.5. 

Having discussed the mechanisms and making ourselves familiar with the magnitudes of the parameters involved in both the 
collectice electric and magnetic modes of exciting neutron production, we shall devote the next three sections to a more detailed description of how it is realized in the three different physical cases: metallic hydride cells, exploding wires and the solar corona.    

\section{3. Metallic Hydride Cells} 

While our discussion would hold for any metallic hydride, we shall concentrate here on palladium hydrides which are particularly suited for our purpose since on such a loaded hydride there will be a full proton layer on the hydride   
surface. On this surface there will then exist coherent proton oscillations. That is, all the protons will be undergoing a collective oscillation known as the surface plasmon mode. Let us determine the size of this plasma frequency \begin{math}\Omega\end{math} and estimate the mean electric field 
\begin{math}\bar{E}\ =\ \sqrt{\bar{{\bf E}^2}}\end{math} generated on the surface.

Suppose a proton of mass \begin{math}M_p\end{math} is embedded in a sphere with a mean electronic charge density \begin{math}\rho_e = (-e) n\end{math}. If the proton suffers a small displacement \begin{math}{\bf u}\end{math}, then an electric field will be created
\begin{equation}
\label{H1}
e {\bf E} = -\big{(}\frac{4 \pi e^2 n}{3} {\bf u} \big{)} = - M_p \Omega^2 {\bf u},
\end{equation}
to satisfy Gauss' law \begin{math}div {\bf E}\ =\ 4\pi \rho_e \end{math}. This electric field will try to push back the proton to the center of the sphere. The equation of motion of the proton 
\begin{math}M_p \ddot{\bf u}\ =\ e {\bf E}\ =\ -\Omega^2 M_p {\bf u}\end{math} yields the oscillation. This equation also furnishes the relationship between the mean electric field and the mean proton displacement
\begin{math}\end{math}
\begin{equation}
\label{H2}
e^2 \bar{{\bf E}^2} = \big{(} \frac{4 \pi e^2 n}{3}\big{)}^2 \bar{{\bf u}^2}.  
\end{equation} 
We may estimate the strength of the mean electric field by taking the mean electron number density at the position of the proton
\begin{equation}
\label{H21}
n \approx\ |\psi(0)|^2\ =\ \frac{1}{\pi a^3}.
\end{equation}  
Defining the atomic electric field at a distance \begin{math}a\end{math} away as
\begin{equation}
\label{H22}
{\cal E}_a = \frac{e}{a^2}\ \approx\ 5.1 \times\ 10^{11} Volts/meter,
\end{equation}
we obtain
\begin{equation}
\label{H23}
\bar{{\bf E}^2} = {\cal E}_a^2 (\frac{16}{9}) \big{(}\frac{\bar{{\bf u}^2}}{a^2}\big{)}
\end{equation}
Neutron scattering experiments on palladium hydride clearly indicate a sharply defined oscillation peak for  \begin{math}(\hbar \Omega/e)\ \approx\ 60\ millivolts\end{math}, as quoted in sec.2. Such a collective proton motion at an infra-red frequency will resonate with electronic surface plasmon oscillations leading to the local breakdown of the Born-Oppenheimer approximation\cite{bo}. They will also lead to large collective proton oscillation amplitudes. This explains the large mean proton displacement \begin{math}\bar{u}\ \approx\ 2.2\ Angstroms\end{math} estimated from the neutron scattering experiments and a mean electric field estimate through Eq(\ref{H23})
\begin{equation}
\label{H3}
\bar{E}\ \approx\ 28.8 \times 10^{11} Volts/meter\,
\end{equation}
which is over 10 times larger than \begin{math}\cal{E}\end{math} thus proving that the plasma oscillations on metallic hydride surfaces do provide internal local electric fields more than sufficient to accelerate the surface plasmon polariton
electrons in overcoming the threshold barrier.

Our description about the acceleration of an electron due to charge oscillations on the surface can be recast in a manifestly Lorentz and gauge covariant form to imply that the free electron mass \begin{math}m\end{math}  has been ``dressed up'' or renormalized to a higher value \begin{math}\tilde{m}\ =\ \beta m\end{math} for the surface electrons (vedi Eq.(\ref{EW2})). Once \begin{math}\beta\ \geq\ \beta_o\ \approx\ 2.53\end{math}, neutron production through weak interaction is kinematically allowed.

For these heavy surface plasmon polariton electrons \begin{math}\tilde{e}^-\end{math} with more than sufficient mass to enable the weak interaction \begin{math}\tilde{e}^-\ +\ p\ \rightarrow\ n\ +\ \nu_e\end{math} to proceed, we turn to a rough order of magnitude estimate of this reaction rate. For this purpose, we may employ (i) the usual Fermi point-like left-handed interaction with coupling constant \begin{math}G_F\end{math}\cite{SM}, (ii) a heavy electron mass \begin{math}\tilde{m}\ =\ \beta m\end{math} and, (iii) the small neutron-proton mass difference \begin{math}\Delta\ =\  (M_n - M_p)\ \approx\ 1.3\ MeV/c^2\end{math}. To make an order of magnitude estimate of the rate of this reaction, we observe that this rate which (in lowest order of perturbation theory) is proportional to \begin{math}G_F^2\end{math} must on dimensional grounds scale with the fifth power of the electron mass. Hence a rough dimensional analysis estimate would give
\begin{equation}
\label{H4}
\Gamma ({\tilde e}^-p\rightarrow\ n\nu_e)\ \approx\ \big{(}\frac{G_F m^2c}{\hbar^3} \big{)}^2 (\frac{mc^2}{\hbar}) (\frac{{\tilde m} - \Delta}{\Delta})^2 
\end{equation}   
Numerically, this would imply a rate
\begin{equation}
\label{H5}
\Gamma ({\tilde e}^-p\rightarrow\ n\nu_e)\ \approx\ 7 \times 10^{-3}\ (\frac{{\tilde m} - \Delta}{\Delta})^2\ Hertz, 
\end{equation} 
which in turn implies
\begin{equation}
\label{H6}
\Gamma ({\tilde e}^-p\rightarrow\ n\nu_e)\ \approx\ 1.2 \times 10^{-3}\ (\beta - \beta_o)^2\ Hertz. 
\end{equation} 
If we assume a surface density of \begin{math}10^{16}/cm^2\end{math} (heavy electron- proton) pairs, we arrive at  the following estimate for the rate of weak neutron production on a hydride surface
\begin{equation}
\label{H7}
{\tilde w}({\tilde e}^-p\rightarrow\ n\nu_e)\ \approx\ (1.2 \times 10^{13}\ /cm^2  /second)\ (\beta - \beta_o)^2, 
\end{equation} 
which is substantial indeed.  

The neutrons so produced will be of ultra low momentum since their production is through the collectively oscillating protons acting in tandem from a patch on the surface. The wavelength of the neutrons may be estimated to be about \begin{math} \lambda\ \approx\ 10^{-3}\ cm\end{math}. Such long wavelength neutrons will get absorbed with an extremely high probability by the nuclei on the surface since the neutron absorption cross-section would be very large.
This can be seen by computing the total neutron cross-section through the optical theorem which relates it to the forward elastic (\begin{math}n-Nucleus\end{math})\ amplitude
\begin{equation}
\label{H9}
\sigma_T(n\ +\ Nucleus\ \rightarrow\ anything) = (\frac{4 \pi}{k})\Im m f(k,{\bf 0}).   
\end{equation}    
Let \begin{math}f(k,{\bf 0})\ =\ a\ +\ i b\end{math}, with \begin{math}b\ \approx\ 1\ fermi\end{math}. Then, we obtain
\begin{equation}
\label{H10}
\sigma_T(n\ +\ Nucleus\  \rightarrow\ anything) = 2\lambda b\ \approx\ 2 \times\ 10^8\ barns,   
\end{equation} 
a very large value. This not only shows a hefty rate for the production of nuclear transmutations through a rapid absorption of neutrons but it also shows that the mean free path of a neutron \begin{math}\Lambda \end{math} is of the order of a few atomic distances. In fact, given the density of neutron absorbers \begin{math}n_{abs}\ \approx\ 10^{22}/cm^3\end{math}, we may estimate
\begin{equation}
\label{H11}
\Lambda^{-1} = n_{abs} \sigma_T;\ \Lambda\ \approx\ 50\ Angstroms 
\end{equation}
Hence, practically all produced neutrons will get absorbed with essentially a zero probability of finding a free neutron.

The observed electromagnetic radiation from the surface heavy electrons will be confined essentially to low energy photons reaching up to soft X-rays with practically no MeV range photons being radiated since the mean free path of any produced gamma rays in the few MeV range would be very short, about a few Angstroms. 

To recapitulate: The surface charge oscillation plasmons provide enough collective energy for the production of heavy mass electrons which in turn lead to the production of low momentum neutrons. Such neutrons get readily absorbed and their production dynamics produces their own neutron and built-in gamma ray ``shields''. The observable end products would just be the nuclear transmutations triggered by the absorbed neutrons. A plethora of nuclear reactions are thereby possible. One such complete nuclear chain cycle with a high Q-value is as follows. Let us assume that the surface is coated with Lithium. Successive absorption of neutrons by Lithium will produce \begin{math}^4_2He\end{math}: 
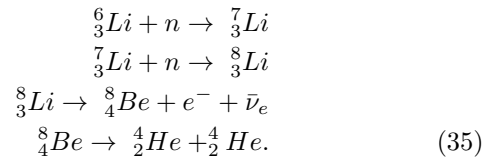
\begin{eqnarray}
\nonumber
^6_3Li + n \rightarrow\ ^7_3Li\\
\nonumber
^7_3Li + n \rightarrow\ ^8_3Li\\
\nonumber
^8_3Li \rightarrow\ ^8_4Be + e^- + \bar{\nu}_e\\
\label{H12}
^8_4Be \rightarrow\ ^4_2He + ^4_2He.
\end{eqnarray}
The heat produced through the above reaction is quite high: \begin{math}Q[^6_3Li\ +\ 2n\ \rightarrow\ 2\ ^4_2He\ + e^- + \bar{\nu}_e]\ \approx\ 26.9\ MeV\end{math}.

On the other hand,  \begin{math}^4_2He\end{math} can successively absorb neutrons and, through the formation of intermediate halo nuclei, reproduce Lithium  
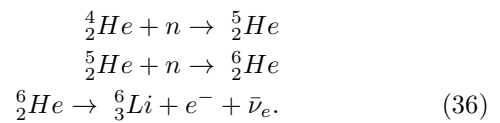
\begin{eqnarray}
\nonumber
^4_2He + n \rightarrow\ ^5_2He\\
\nonumber
^5_2He + n \rightarrow\ ^6_2He\\
\label{H13}
^6_2He \rightarrow\ ^6_3Li + e^- + \bar{\nu}_e.
\end{eqnarray}
The heat from the reaction in Eq.(\ref{H13}) is \begin{math}Q[^4_2He\ +\ 2n\ \rightarrow\  ^6_3Li\ + e^- + \bar{\nu}_e]\ \approx\ 2.95\ MeV\end{math}. The complete nuclear cycle as described in Eqs.(\ref{H12}) and (\ref{H13}) taken together would release a substantial total heat through nuclear transmutations. Other Lithium initiated processes would produce both \begin{math}^4He\end{math} and \begin{math}^3He\end{math}.

\section{4. Exploding Wires}

In sec.2, we have outlined our explanation of nuclear transmutations and fast neutrons which have been observed to emerge when large electrical current pulses passing through wire filaments are induced to explode. If a strong current pulse, large on the scale of \begin{math} I_0  \end{math}, defined in Eq.(\ref{W3}), passes through a thin wire filament, then the magnetic field which loops around the current, exerts a very large Maxwell pressure on surface area elements, compressing, twisting and pushing into the wire. When the magnetic Maxwell pressure grows beyond the tensile strength of the wire material at the hot filament temperature, the wire first expands, then begins to melt and finally disintegrates. There now exist slow-motion pictures which verify that indeed the wire expands, melts and disintegrates. All of this is readily understood. If the heating rate is sufficiently fast, then the hot wire may emit thermal radiation at a very high noise temperature. The thermal radiation from exploding Tungsten filaments exhibits X-ray frequencies indicating very high electron kinetic energies within the filament. Due to the electron kinetic pressure, the wire diameter starts to increase yielding a filament dense gas phase but still with some liquid droplets. The final explosive product consists of a hot plasma colloid containing some small dust particles of the original wire material. These products cool off into a gas and some smoke as is usual for explosions.

As discussed in sec.2, we want to understand how LENRs can be initiated in an exploding wire current pulse with a strong current (with its peak value substantially higher than \begin{math}I_o \end{math}) produced by a capacitor discharge with an initial voltage of only \begin{math}30\ KeV \end{math}\cite{Wendt1,Wendt2}. We also want to understand, by contrast, why when Rutherford had fired a much higher energy \begin{math}100\ KeV\end{math} but dilute beam of electrons into a Tungsten target he did not observe any nuclear reactions\cite{Rutherford}.

A typical electron in the current with a mean kinetic energy \begin{math}15\ KeV \end{math} would have an average speed \begin{math}(v/c)\ \approx\ 0.25 \end{math}. On the other hand, even for such low mean speed, the chemical potential given in Eq.(\ref{EW8}), for \begin{math}(I/I_o)\ \approx\ 200\end{math} becomes large
\begin{equation}
\label{E1}
\mu\ \approx\ (mc^2)(200)(0.25)\ =\  25\ MeV,
\end{equation}
comfortably sufficient for an electron to induce a weak interaction LENR. Overall energy conservation would of course require that only a certain fraction about \begin{math}(15\ KeV/25 MeV)\ =\ 6\times\ 10^{-4}\end{math} of the total number of electrons in the current would be kinematically allowed to undergo weak interactions.

Let us now briefly discuss why Rutherford with his much higher energy -but dilute- beam of electrons did not observe any nuclear reactions. The reason is rather simple. In the vacuum, there is a mutual Coulomb repulsion between the electrons in the beam which compensates the mutual Amperian current attraction. In the exploding wire filament, on the other hand, the repulsive Coulomb part is screened by the background positive charge but leaves intact the Amperian current attraction thereby allowing the possibility of nuclear reactions.  

\section{5. Solar Corona and Flares}

As stated in sec.2, oppositely directed Amperian currents of electrons and protons loop around the walls of a magnetic flux tube which exits out of one sun spot into the solar corona to enter back into another sun spot. The magnetic flux tube is held up by magnetic buoyancy. We consider here the dynamics of how very energetic particles are produced in the solar corona and how they induce nuclear reactions well beyond the solar photosphere. Our explanation, centered around Faraday's law, produces the notion of a solar accelerator very similar to a betatron\cite{K1,K2}. A betatron is a step-up transformer whose secondary coil is a toroidal ring
of particles circulating around a time-varying Faraday flux tube.

We may view the solar flux tube to act as a step-up transformer which passes some circulating particle kinetic energy from the solar photosphere outward to other circulating particles in the solar corona. The circulating currents within the photosphere are to be considered as a net current \begin{math}I_{{\cal P}}\end{math} around a primary coil and the circulating currents high up in the corona as a net current \begin{math}I_{{\cal S}}\end{math}. If \begin{math}K_{{\cal P}}\end{math} and \begin{math}K_{{\cal S}}\end{math} represent the kinetic energies, respectively, in the primary and the secondary coils, the step up transformer power equation reads
\begin{equation}
\label{C1}
|{\dot K}_{{\cal P}}| = |V_{{\cal P}}I_{{\cal P}}| = |V_{{\cal S}}I_{{\cal S}}| = |{\dot K}_{{\cal S}}|,
\end{equation}
where \begin{math}V_{{P}}\end{math} and \begin{math}V_{{S}}\end{math} represent the voltages across the primary and the secondary coils, respectively. The total kinetic energy transfer reads
\begin{equation}
\label{C2}
\Delta K_{{\cal P}} = \int (dt)|V_{{\cal P}}I_{{\cal P}}| =\int (dt) |V_{{\cal S}}I_{{\cal S}}| = \Delta K_{{\cal S}}.
\end{equation}
In essence, what the step up transformer mechanism does is to transfer the kinetic energy distributed amongst a very large number of charged particles in the photosphere -via the magnetic flux tube- into a distant much smaller number of charged particles located in the solar corona, so that a small accelerating voltage in the primary coil produces a large accelerating voltage in the secondary coil. The transfer of kinetic energy is {\it collective} from a larger group of particles into a smaller group of particles resulting in the kinetic energy per charged particle of the dilute gas in the corona becoming higher than the kinetic energy per particle of the more dense fluid in the photosphere.

We may convert the above into a temperature - kinetic energy relationship by saying that the temperature of the dilute corona will be much higher than the temperature of the more dense fluid photosphere. If and when the kinetic energy of the circulating currents in a part of the floating flux tube becomes sufficiently high, the flux tube would become unstable and explode into a solar flare which may be accompanied by a coronal mass ejection. There is a rapid conversion of the magnetic energy into charged particle kinetic energy. These high energy products from the explosion initiate nuclear as well as elementary particle interactions, some of which have been detected in laboratories on the earth.

Recent NASA and ESA pictures show that the surface of the sun is covered by a carpet-like interwoven mesh of magnetic flux tubes of smaller dimensions. Some of these smaller structures possess enough magnetic energy to lead to LENRs through a continual conversion of their energy into particle kinetic energy. Occurrence of such nuclear processes in a roughly steady state would account for the solar corona remaining much hotter than the photosphere. Needless to say that our picture belies the notion that all nuclear reactions are contained within the core of the sun. On the contrary, it provides strong theoretical support for experimental anomalies\cite{Fowler}\cite{Wurz} such as short-lived isotopes\cite{Cowley}\cite{Goriely}\cite{Lodders} that have been observed
in the spectra of stars having unusually high average magnetic fields.

For the transformer mechanism to be fully operational in the corona, the coronal electrical conductivity must not be too large. Useful experimental bounds on an upper limit to this conductivity may be obtained through its effect on measurements of gravitational bending of light near the sun as it traverses the solar corona. Successful measurements of the gravitational bending of electromagnetic waves with frequencies
in the visible and all the way down to the high end of the radio spectrum are legion. These experiments provide a direct proof that any coronal conductivity disturbance on the expected gravitational bending of electromagnetic waves for frequencies down to \begin{math}12.5\ GigaHertz\end{math} must be negligible. Error estimates from even lower frequency radio wave probes used for gravitational bending\cite{Weinberg}
\cite{Muhleman} put the coronal conductivity in the megahertz range. For comparison, we note that the typical conductivity of a good metal would be more than ten orders of magnitude higher\cite{salt}. The corona is close to being an insulator and eons away from being a metal and their is no impedimenta toward sustaining electrical fields within it\cite{Foukal,Feldman,Stas}. Thus, our proposed transformer mechanism and its subsequent predictions for the corona remain intact.  

The spectacular solar flare, which occurred on July 14, 2000 and the measurement of the excess muon flux associated with this flare by the CERN L3+C group\cite{C9} offered a unique opportunity to infer that protons
of energies in excess of $40$ GeV were produced in the solar corona. Likewise, the BAKSAN underground muon
measurements\cite{Baksan} provided evidence for protons of energies in excess of $500$ GeV in the solar flare of September 29, 1989. The very existence of primary protons in this high energy range provides strong evidence for the numbers provided in Eq.(\ref{EW18}). Hence, for large solar flares in the corona, electrons and protons must have been accelerated well beyond anything contemplated by the standard solar model. This in turn provides the most compelling evidence for the presence of large scale electric fields and the transformer or betatron mechanism since we do not know of any other process that could accelerate charged particles to beyond even a few GeV let alone hundreds of GeVs.

\section{6. Summary and Concluding Remarks} 

We may summarize by saying that three seemingly diverse physical phenomena viz., metallic hydride cells, exploding wires and the solar corona do have a unifying theme. Under appropriate conditions which we have now well delineated, in all these processes electromagnetic energy gets collectively harnessed to provide enough kinetic energy to a certain fraction of the electrons to combine with protons (or any other ions present) and produce neutrons through weak interactions. The produced neutrons then combine with other nuclei to induce low energy nuclear reactions and transmutations. Lest it escape notice let us remind the reader that all three interactions of the standard model (electromagnetic, weak and nuclear) are essential for an understanding of these phenomena. Collective effects, but no new physics for the acceleration of electrons beyond the Standard Model needs to be invoked. However, we have seen that certain paradigm shifts are necessary. On the surface of a metallic hydride cell with surface plasmon polariton modes, protons collectively oscillate along with the electrons. Hence, the Born-Oppenheimer approximation (which assumes that the proton is rigidly fixed) breaks down and should not be employed. Similarly, in the solar corona, the electronic density and the electrical conductivity are sufficiently low. Hence there is not much charge screening of the  electric fields involved. Strong electric fields generated by time-dependent magnetic fields through Farday's laws are sustained in the corona and the betatron (or transformer) mechanism remains functional. Were it not so, electrons and protons could not have been accelerated to hundreds of GeV's and there would have been no production of high energy muons, certainly not copious enough to have reached Earth in sufficient numbers to have been observed by the L3+C collaboration at LEP\cite{C9} or by the BAKSAN underground laboratory\cite{Baksan}. We are unaware of any other alternative scheme for obtaining this result. The betatron mechanism also naturally explains a variety of observed experimental results such as unexpected nuclear transmutations and high energy cosmic rays from the exterior of the sun or any other astronomical object endowed with strong enough magnetic activity such as active galactic nuclei.  

The analysis presented in this paper leads us to conclude that realistic possibilities exist for designing LENR devices capable of producing ``green energy'', that is production of excess heat at low cost without lethal nuclear waste, dangerous gamma rays or unwanted neutrons. The necessary tools and the essential theoretical know-how to manufacture such devices appear to be well within the reach of presently available technology. Vigorous efforts must now be made to develop such devices whose functionality requires all three interactions of the Standard Model acting in concert.  
       
\section{Acknowledgements} 
Over the past few years, various phases of this work  have been presented by us in seminars and lectures at the Universities of Milan, Perugia, Rome I \& III; Olivetti Research Center in Milan; INFN Italian National Laboratory at Frascati; Inter University Accelerator Center in New Delhi, India and various US government departments and agencies in Washington DC USA. We take this opportunity to thank our colleagues for inviting us to their institutions and for many interesting and useful remarks.


\begin{thebibliography}{08}
\bibitem{mom} Under the same electric field, the protons will gain an equal and opposite momentum
to that of the electrons, hence their CM would stay at rest (if we neglect their very small initial momenta).
\bibitem{bo} Let us recall that in the Born-Oppenheimer approximation, the nuclear coordinates are assumed fixed, only the electron coordinates are assumed dynamical. But on the thin surface of a mono layer, where collective plasmon oscillation modes for both the electrons and the protons exist, the approximation becomes untenable. In the bulk of the metal, the situation is different due to the screening of charges. The acceleration of electrons and the resultant neutron production being discussed here is all on the mono layer surface not in the bulk.
\bibitem{SM} In the Standard Model of particle physics, this process is described through the charged  \begin{math}W\end{math}-boson exchange whose mass is \begin{math}M_W\ \approx\ 80\ GeV/c^2\end{math}, corresponding to a rest energy so large on the scale of energies being discussed here, that the Fermi point-like limit of the weak interaction is an excellent approximation, with the identification \begin{math}G_F/(\hbar c)^3\ =\ (\pi \alpha/\sqrt{2})/(M_W c^2 sin\theta_W)^2\ \approx\ 1.1 \times 10^{-5}\ GeV^{-2}\end{math}. Here enters also the weak angle through \begin{math}sin^2\theta_W\ \approx\ 0.23\end{math}.
\bibitem{salt} Salty ocean water has a conductivity of about \begin{math}1.5\ \times\ 10^{11}Hertz\end{math}.
\end{thebibliography}
\end{document}